\documentstyle[12pt]{article}
\begin{document}

\begin{flushright}
{\sf Portmouth University\\
Relativity and Cosmology Group\\
{\em Preprint} RCG 95/13}
\end{flushright}
\[ \]
\begin{center}
{\large\bf Anisotropy and inhomogeneity of the universe from
$\Delta T/T$}
\[ \]
Roy Maartens\footnote{School of Mathematical Studies,
Portsmouth University, Portsmouth PO1 2EG, England\\
Member of Centre for
Nonlinear Studies, Witwatersrand University, South Africa},
George F.R. Ellis\footnote{Department of Mathematics and
Applied Mathematics,
University of Cape Town, Cape Town 7700, South Africa} and
William R. Stoeger\footnote{Vatican Observatory Research
Group, Steward Observatory, University of Arizona, Tucson
AZ 85721, USA}
\[ \]
\[ \]
{\bf Abstract}
\end{center}
\[ \]
A recent paper (Martinez--Gonzalez \& Sanz 1995) showed that if
the universe is homogeneous but anisotropic, then the small
quadrupole anisotropy in the cosmic microwave background radiation
implies that the spacetime anisotropy is very small. We point out that
more general results may be established, without assuming
a priori homogeneity. We have proved that small
anisotropies in the microwave background imply
that the universe is almost Friedmann--Robertson--Walker.
Furthermore, the quadrupole and octopole place direct and
explicit limits on the degree of anisotropy and inhomogeneity, as
measured by the shear, vorticity, Weyl
tensor and density gradients. In the presence of inhomogeneity, it is
only possible to set a much weaker limit on the shear than that
given by Martinez--Gonzalez \& Sanz.
\newpage

Martinez--Gonzalez \& Sanz (1995) (hereafter MS)
point out that part of the
foundation for the standard Friedmann--Robertson--Walker (FRW)
model of the universe is to prove that the small anisotropies in
the cosmic microwave background radiation (CBR)
imply that only small deviations from homogeneity and
isotropy of the universe are possible. They prove a particular
special case of such a general theorem: {\it if the universe is
homogeneous and flat (i.e. a Bianchi I model), and if the
dynamical effects of radiation are neglected,
then the small quadrupole moment
of the CBR implies that the anisotropy of the universe
(i.e. deviation from an FRW model) is very small.}

In fact, the general theorem has been proved by Stoeger et al.
(1995). This theorem generalises the exact--isotropy theorem of
Ehlers et al. (1968) to the case of almost--isotropy.
It follows without making assumptions about the spacetime
inhomogeneity and anisotropy, and without neglecting the dynamical
effect of radiation:\\

\noindent{\sf Theorem:}
{\it if all fundamental observers measure the CBR to be almost
isotropic in an expanding universe, then that universe is locally
almost spatially homogeneous and isotropic
(i.e. it is almost FRW) after last scattering.}\\

This result provides a consistent theoretical
foundation for the standard analyses of the CBR based on the
Sachs--Wolfe effect (see e.g. Hu \& Sugiyama 1995),
which {\it assume} that the universe is almost FRW.
Note that the theorem
incorporates the Copernican Principle, i.e.
if the CBR is almost isotropic for our galaxy, then
it is almost isotropic for {\it all} galaxies,
since we do not occupy a privileged position.\footnote{The
Copernican
principle is in fact partially testable via the Sunyaev--Zeldovich
effect (Maartens et al. 1995b; Goodman 1995).}

The proof of the theorem is based on a covariant and gauge--invariant
analysis of the Einstein--Boltzmann equations governing dust and
radiation after last scattering. This formalism
is then applied by Maartens et al. (1995a,b) (hereafter MESa,b)
to a quantitative
investigation of the relationship between temperature anisotropies
and the inhomogeneity and anisotropy of the universe. Before we
describe the general limits on spacetime
inhomogeneity and anisotropy that
are imposed by CBR anisotropies, we situate the special result
of MS within the general results of MESa,b.

MS make the non--observational assumption that the universe
has exact Bianchi I symmetry. Strictly, this rules out density
perturbations, vorticity and gravitational wave perturbations,
and also
excludes the cases where the FRW background has non--critical
matter density (i.e. is not flat).
However, the model of MS is clearly intended as a first step
towards the general case. As such, we can provide
an observational basis for their model
via one of the results of MESa (p. 1532):\\
{\it if the residual dipole of the CBR temperature anisotropy
vanishes to first order, and if
the quadrupole and octopole are spatially
homogeneous to first order, then
the spacetime is locally Bianchi I to first order.}\\

Indeed this special case is still more general than the MS model,
since the spacetime is not exactly Bianchi I, but only to first order.
With this qualification,
the result of MS may be interpreted as
the special case of the general theorem of Stoeger et al. (1995)
which applies if the quadrupole and octopole of the temperature
anisotropy are almost spatially homogeneous, and if the residual
dipole vanishes to first order.

By using exact solutions for Bianchi I dust models (and therefore
ignoring the radiation energy density after last scattering),
MS deduce the following limit imposed by COBE observations
on the relative distortion at the current time:
\begin{equation}
\left({\sigma\over\Theta}\right)_0 \leq 6.9\times 10^{-10}\,,
\label{1}
\end{equation}
where $\sigma$ is the shear anisotropy and
$\Theta(=3H)$ is the rate of expansion. We can derive an
independent confirmation of the magnitude of this limit.
In MESa (Eq. (99)), we showed that
if the residual dipole, quadrupole and octopole
are spatially homogeneous
to first order (and without any further assumptions), then
\begin{equation}
\left({\sigma\over\Theta}\right)_0<\left({16\Omega_\gamma\over 15\Omega}
\right)_0\epsilon_2\,,
\label{2}
\end{equation}
where $\Omega_\gamma$, $\Omega$ are the density parameters for radiation
and matter respectively, and $\epsilon_2$ is the upper limit on the
quadrupole. For the large--scale anisotropies probed by COBE,
we follow MS and take
$\epsilon_2 \approx 10^{-5}$. In the case of critical matter
density, which is also assumed by MS, one has
(Kolb \& Turner 1990, p. 503)
\begin{equation}
\left({\Omega_\gamma\over\Omega}\right)_0\approx
2.5h^{-2}\times 10^{-5}\,,
\label{2a}
\end{equation}
where the Hubble constant is given by $H_0=100h$ km/s/Mpc, and
$0.4\leq h\leq 1$. Then Eq. (\ref{2}) and Eq. (\ref{2a}) give
\begin{equation}
\left({\sigma\over\Theta}\right)_0< 2.7h^{-2}\times 10^{-10}\,,
\label{3}
\end{equation}
which is consistent with Eq. (\ref{1}). Furthermore, the
electric Weyl tensor $E_{ab}$
(not discussed by MS) is bounded by
\begin{eqnarray}
\left(\sqrt{E_{ab}E^{ab}}\right)_0&<&{\textstyle{12\over5}}\left(
\Omega_\gamma\right)_0c^{-2}H_0^2\epsilon_2 \nonumber\\
&\approx& {\textstyle{20\over3}}\times 10^{-17}~{\rm per~(Mpc)}^2\,.
\label{4}
\end{eqnarray}

Equations (\ref{3}) and (\ref{4}) are sufficient to characterise
the small deviation from isotropy of a spacetime with nearly Bianchi
I symmetry. But in general, anisotropy and inhomogeneity of
the universe are determined not only by $\sigma$
and $E_{ab}$, but also by
$\omega$ (vorticity), $H_{ab}$ (magnetic Weyl tensor), and by
the spatial gradients of $\rho$ (matter density),
$\Theta$ and $\mu$ (radiation energy density).

In MESa,b we derived limits on all these quantities, explicitly
in terms of upper bounds on the multipoles of the CBR temperature
anisotropy. These limits do not depend on assuming inflationary or
other models for the source of perturbations. It turns out that to
first order, only the first three multipoles - the residual dipole,
the quadrupole and the octopole - enter the Einstein--Boltzmann
equations. Thus only the
first three multipoles play a direct role in limiting the covariant
quantities that measure inhomogeneity and anisotropy.

For
example, the relative distortion in general is limited by (MESb,
Eq. (24)):
\begin{equation}
\left({\sigma\over\Theta}\right)_0< {\textstyle{8\over3}}\epsilon_2
+\epsilon_2^*+5\epsilon_1^{\dagger}+{\textstyle{9\over7}}
\epsilon_3^{\dagger}\,,
\label{5}
\end{equation}
where $\epsilon_2^*\Theta_0$ is the bound on the
time rate of change of
the quadrupole, $\epsilon_1^{\dagger}\Theta_0/c$ is the bound on the
spatial
gradient of the residual dipole, and $\epsilon_3^{\dagger}\Theta_0/c$
is the bound on the spatial gradient of the octopole.

In the special
case of spatially homogeneous multipoles, we have
$\epsilon_1^\dagger=0=\epsilon_3^\dagger$ in Eq. (\ref{5}).
However, it is possible to obtain a much tighter limit. This
follows since
the Einstein--Boltzmann equations may be decoupled into independent
evolution and constraint equations. The evolution equations may be
reduced to a third order ordinary differential equation in the
shear (MESa, Eq. (80)). The electric Weyl tensor and other
quantities are then determined in terms of the shear. In this
way, the severe limit of Eq. (\ref{2}) on the shear
is obtained, leading to
the confirmation of the MS result. However, in the
general inhomogeneous case, no decoupling
or integration is possible, and one has to deal with the full coupled
system of equations. This leads to the much weaker shear limit of Eq.
(\ref{5}).

The general limits are complicated by the bounds not only on
the multipoles of the CBR temperature, but also on their derivatives.
These derivatives are not directly accessible to current observations.
We need to estimate the bounds on the derivatives of the
multipoles in terms of the bounds on the multipoles themselves,
which are accessible to observations. We make the reasonable
assumptions that:
\begin{itemize}
\item {\bf (a)} the spatial gradients are not greater than the
time derivatives;\\
\item {\bf (b)} the time derivative of a multipole may be estimated as
the multipole divided by
the characteristic time--scale $T/|\dot{T}|$ of the CBR.
\end{itemize}

With these assumptions, the relative distortion limit of Eq. (\ref{5})
becomes
\begin{equation}
\left({\sigma\over\Theta}\right)_0< {\textstyle{5\over3}}\epsilon_1
+3\epsilon_2+{\textstyle{3\over7}}\epsilon_3\,,
\label{6}
\end{equation}
where $\epsilon_1,\epsilon_2,\epsilon_3$ are the bounds on the
residual dipole, quadrupole and octopole respectively. The remaining
limits are given in MESb (Eq. (30)--(36)).

It is usually assumed that the residual dipole (i.e. after correction
for local peculiar velocity) is negligible, although this does not
follow from CBR observations (Copeland et al. 1993). We will adopt
this standard assumption. It is also reasonable to neglect the
radiation energy density at the current time. Thus we have the
additional assumption:
\begin{itemize}
\item {\bf (c)}$\quad\left(\Omega_\gamma\right)_0\ll \Omega_0\,,~~
\epsilon_1\ll~{\rm max}~(\epsilon_2,\epsilon_3)=\alpha
\times 10^{-5}\,,$
\end{itemize}

\noindent where $\alpha$ is determined by observations.
For the large scales
probed by COBE, $\alpha$ is of the order of 1.
Given assumptions {\bf (a--c)}, we can use the results of MESa,b
to compute the limits on the present size of all the (covariant
and gauge--invariant) quantities that determine the inhomogeneity
and anisotropy of the universe:
\begin{equation}
\left({{\sigma} \over \Theta}\right)_0 < 4\alpha\times 10^{-5}\,,~~
\left({{\omega} \over \Theta}\right)_0 < \alpha\times10^{-5} \,,
\label{7}
\end{equation}
\begin{equation}
\left(\sqrt{E_{ab}E^{ab}}\right)_0 < [{\textstyle{19\over5}}
+{\textstyle{2\over15}}\Omega_0]\alpha h^2\times 10^{-12}
{}~{\rm per~}({\rm Mpc})^2\,,
\label{8}
\end{equation}
\begin{equation}
\left(\sqrt{H_{ab}H^{ab}}\right)_0 < 4\alpha h^2\times 10^{-13}
{}~{\rm per~}({\rm Mpc})^2\,,
\label{9}
\end{equation}
\begin{equation}
\left({{|\vec{\nabla}\mu|} \over \mu}\right)_0 <
{\textstyle{5\over3}}\alpha h\times 10^{-9}~{\rm per~Mpc}\,,
\label{10}
\end{equation}
\begin{equation}
\left({|\vec{\nabla}\Theta| \over \Theta}\right)_0 <
[{\textstyle{8\over3}}+{\textstyle{4\over3}}\Omega_0]\alpha h
\times 10^{-9}~{\rm per~Mpc}\,,
\label{11}
\end{equation}
\begin{equation}
\left({{|\vec{\nabla}\rho|} \over \rho}\right)_0 < \left[
{\textstyle{14 \over 3}}\left(\Omega_0\right)^{-1}
+{\textstyle{1\over6}}\right]\alpha h
\times 10^{-8}~{\rm per~Mpc}\,.
\label{12}
\end{equation}

Comparing the general result Eq. (\ref{7}) with the Bianchi I result
Eq. (\ref{1}) of MS, we see that the limit on the relative distortion
is significantly weaker - by about 5 orders of magnitude -
when inhomogeneity is present. The limit on the relative
vorticity is comparable to that on distortion. In the Bianchi I
model of MS, the vorticity is of course assumed a priori
to be exactly zero.

The magnetic Weyl tensor $H_{ab}$, which is non--zero in the
presence of gravitational waves, also vanishes in the MS case
by assumption. In the general case,
Eq. (\ref{9}) places limits on the presence of
long--wavelength gravitational perturbations.

The smallness of the limit on the density gradient Eq. (\ref{12})
reflects the large scales that are being probed, i.e. about
$10^2$--$10^3$ Mpc. On the scale of the observable universe, i.e.
approximately $3000h^{-1}$ Mpc (Kolb and Turner 1990),
Eq. (\ref{12}) implies
the following approximate limit on the average density contrast:
\begin{equation}
\left({\Delta\rho\over\rho}\right)_0 <
\left[{\textstyle{7\over 5}}\left(\Omega_0\right)^{-1}
+{\textstyle{1\over20}}\right]\alpha\times 10^{-4}\,.
\label{13}
\end{equation}
This is consistent with the
value $\Delta\rho/\rho\sim 10^{-5}$ typically taken to hold at last
scattering (Kolb and Turner 1990).
\[ \]
\[ \]
{\bf Acknowledgements:}
This work was supported by research grants
from the FRD (South Africa). RM was also supported by a research grant
from Portsmouth University.
\[ \]

\end{document}